\PassOptionsToPackage{pdfpagelabels=false}{hyperref} 
\documentclass[usenatbib,fleqn]{mnras}
\usepackage{fixltx2e,longtable,ifpdf,amsmath,amssymb,wasysym}
\usepackage{times,nicefrac,bm}
\ifpdf
  \usepackage{aeguill}
  \usepackage[pdftex]{graphicx,color}
\else
  \usepackage[dvips]{graphicx,color}
\fi

\newcommand{\teff}{\ensuremath{T_{\rm eff}}}
\newcommand{\teffl}{\ensuremath{T_{\rm eff(\ell)}}}
\newcommand{\logg}{\ensuremath{\log{(g)}}}

\newcommand{\msun}{\ensuremath{\mbox{M}_{\odot}}}
\newcommand{\rsun}{\ensuremath{\mbox{R}_{\odot}}}

\newcommand{\mjup}{\ensuremath{\mbox{M}_{\jupiter}}}

\newcommand{\vesini}{\ensuremath{v_{\rm e}\sin{i_{\rm s}}}}
\newcommand{\veq}{\ensuremath{v_{\rm e}}}

\newcommand{\ptfo}{\ensuremath{\mbox{PTFO 8-8695}}}
\newcommand{\ptfob}{\ensuremath{\mbox{PTFO 8-8695b}}}
\newcommand{\exobush}{\ensuremath{\mbox{\textsc{exoBush}}}}

\newcommand{\pz} {\ensuremath{\phantom{0}}}

\newcommand{\kms} {\ensuremath{\mbox{km}\;\mbox{s}^{-1}}}

\newcommand{\bigL}  {\ensuremath{\mathbf{L}}}
\newcommand{\bigo}  {\ensuremath{\mathbf{o}}}
\newcommand{\bigs}  {\ensuremath{\mathbf{s}}}
\newcommand{\omomc} {\ensuremath{\omega/\omega_{\rm c}}}
\newcommand{\dc}{\ensuremath{\mbox{d}}}
\newcommand{\dd}{\ensuremath{\,\mbox{d}}}

\def\fm{\hbox{$.\!\!^{\rm m}$}}

\def\utw{\smash{\rlap{\lower5pt\hbox{$\sim$}}}}
\def\udtw{\smash{\rlap{\lower6pt\hbox{$\approx$}}}}

\title[PTFO 8-8695] {A reappraisal of parameters for the putative
  planet \mbox{PTFO 8-8695b} and its potentially precessing parent
  star.}

\author[Ian D. Howarth ] {Ian D. Howarth\thanks{E-mail:
    i.howarth@ucl.ac.uk} \\ Dept. of Physics and Astronomy, University
  College London, Gower Street, London WC1E 6BT, UK }
\begin{document}

\date{Accepted 2016 January 27. Received 2016 January 23; in original
  form 
2015 November 16}


\maketitle

\label{firstpage}
\begin{abstract}
  Published photometry of fading events in the \ptfo\ system is
  modelled using improved treatments of stellar geometry, surface
  intensities, and, particularly, gravity darkening, with a view to
  testing the planetary-transit hypothesis.  Variability in the
  morphology of fading events can be reproduced by adopting
  convective-envelope gravity darkening, but near-critical stellar
  rotation is required.  This leads to inconsistencies with
  spectroscopic observations; the model also predicts substantial
  photometric variability associated with stellar precession, contrary
  to observations.  Furthermore, the empirical ratio of orbital to
  rotational angular momenta is at odds with physically plausible
  values. An exoplanet transiting a precessing, gravity-darkened star
  may not be the correct explanation of periodic fading events in this
  system.
\end{abstract}


\section{Introduction}

Although the number of confirmed exoplanets is now in the thousands,
the discovery by \citet{vanEyken12} of a possible hot Jupiter
transiting the M-dwarf T-Tauri star \ptfo\ is of particular interest.
Not only is the system exceptionally young ($\sim$3~Myr;
\citealt{Briceno05}), which is of significance in the context of
timescales for planetary formation and orbital evolution, but also it
exhibits variability and asymmetry in the transit light-curves
observed in the two seasons of the Palomar Transient Factory Orion
project (PTFO; \citealt{vanEyken11}).  While part of the
variability may arise through intrinsic stellar effects (such as
starspots), \citet{Barnes13} offered an insightful and credible interpretation that
requires precession of the orbital and rotational
angular-momentum vectors on short timescales ($\sim10^2$~d, to account
for the variable transit depth) coupled with a significantly
gravity-darkened primary (to generate the light-curve asymmetry).

\citet{Barnes13} constructed a detailed numerical realisation of this
model, including periodic precession, which
reproduced the variable light-curve extremely well.  Because of their
interest in physically modelling the precession, they constrained
their model fits by adopting specific values for the stellar mass; and
in order to reduce the number of free parameters they assumed (with
some observational justification) synchronous rotation.
\citet{Kamiaka15} relaxed this assumption, and showed that,
while the system geometry at the two observed epochs is reasonably
well determined, multiple plausible solutions of the intervening
precessional motion exist (as had been anticipated by
\citeauthor{Barnes13}).   Both the \citeauthor{Barnes13} 
and the \citeauthor{Kamiaka15} models predict
that, as observed, transits should not occur at some epochs, as
a consequence of orbital precession.

\citet{Ciardi15} have recently published follow-up observations which
demonstrate further transit-like features in the light-curve
with the correct orbital phasing, albeit at epochs not consistent with the
specific precession model advanced by \citet{Barnes13};
but while this paper
was being prepared,
\citet{Yu15} reported additional observations which challenge the 
\citeauthor{Barnes13} framework.
Thus \ptfob\ remains, at best, only a
\textit{candidate} planet.  
The purpose of the present note is to examine this issue
through more-detailed modelling of the stellar emission, to test, in
particular, the gravity-darkening hypothesis.

\section{Model}

\subsection{Motivation}

Both \citet{Barnes13} and \citet{Kamiaka15} adopted classical
\citet{vonZeipel24} gravity darkening, in which the emergent flux is
locally proportional to gravity; that is,
\begin{equation*}
\teffl \propto g^{\beta}
\end{equation*}
with $\beta=0.25$ (where \teffl\ is to be understood as the
\textit{local} effective temperature).  

However, \citeauthor{vonZeipel24}'s derivation was based on
consideration of a barotropic envelope in which energy transport is
diffusive -- i.e., radiative.  \citet{Lucy67} argued that for stars
with convective envelopes 
the
gravity-darkening exponent $\beta$ is expected to be considerably
smaller;  this argument certainly applies in the case of
\ptfo\ 
(spectral type
$\sim$M3; \citealt{Briceno05}).

Recent work suggests that von~Zeipel's `law' may
overestimate gravity darkening even in radiative envelopes
\citep{Espinosa11};  
and,
while it may be argued that, in respect of
convective envelopes, ``nothing is clear'' \citep{Rieutord15}, it is
surely the case that the gravity-darkening exponent will be less than
in radiative envelopes. The
limited empirical evidence is broadly consistent with
\citeauthor{Lucy67}'s estimate of $\beta \simeq 0.08$ (e.g.,
\citealt{Rafert80, Pantazis98, Djurasevic03, Djurasevic06}), and it is
this value that will be adopted here.

\goodbreak
The  best-fit parameters derived by
\citeauthor{Barnes13} (\citeyear{Barnes13}; their
Table~2)\footnote{`The' radius tabulated therein is the equatorial value
(Barnes, personal communication).}  imply a
ratio of rotational angular velocity to the critical, or break-up,
value of $\omomc \simeq 0.70 \pm 0.04$, and thence
an equatorial:polar temperature ratio of $\sim{0.90\pm 0.02}$.   To
achieve the same temperature contrast with $\beta = 0.08$ (and hence
to achieve  roughly the same degree of light-curve asymmetries) requires
significantly more rapid rotation:
$\omomc \simeq 0.95 \pm 0.02$.  Of course, any change in
\omomc\ leads to changes in the shape of the star (and has
implications for
the rotation period and
the projected equatorial rotation velocity, \vesini), so it is not necessarily obvious
that a consistent solution to the light-curves can be achieved with a
more plausible gravity-darkening exponent.

\subsection{Implementation}
To examine this issue, a modified version of the code for spectrum
synthesis of rapidly rotating stars described by \citet{Howarth01} has
been used.  The code, \exobush,\footnote{The etymology may be
  elucidated by an internet search for `a thousand points of light',
  which is, conceptually, how the tiling of the stellar surface is
  performed.} simply divides the rotationally distorted stellar surface into
a large number of facets; evaluates the
local temperature and gravity at each point; and sums the intensities
$I(\lambda, \mu, T, g)$\footnote{Here $\mu = \cos\theta$, where
$\theta$ is the angle between the surface normal and the line of sight.} to produce a predicted flux, taking into
account occultation by an opaque, nonluminous transiting body of assumed
circular cross-section (e.g., an exoplanet).

\begin{figure}
\begin{center}
\includegraphics[angle=270, scale=0.6]{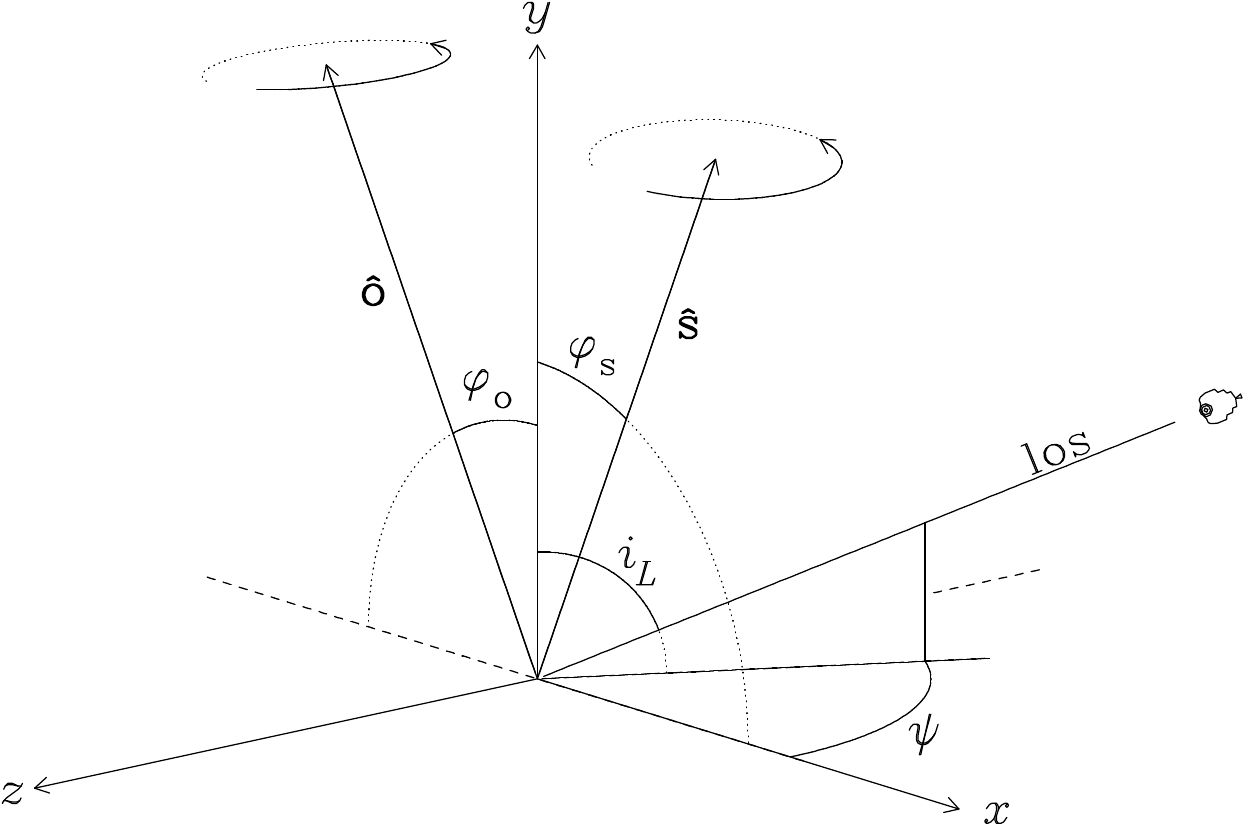} 
\caption{Spin-orbit geometry; 
the $y$ axis coincides with ${\mathbf{L}}$, the sum of
the orbital and stellar-rotation angular momenta
(unit vectors ${\hat{\mathbf{o}}}$,
${\hat{\mathbf{s}}}$), which lie in the $xy$ plane.
The line of sight, `los', is shown for
precessional phase $\psi$;  it is offset from the total angular-momentum
vector ${\mathbf{L}}$ by the angle $i_L$.
The observer's-frame
inclination angles $i_{\rm o}$ 
(between the line of sight and 
${\hat{\mathbf{o}}}$) and
$i_{\rm s}$ 
(between the line of sight and 
${\hat{\mathbf{s}}}$) 
are not shown explicitly.}
\label{fig1}
\end{center}
\end{figure}

\subsubsection{Spin-orbit geometry}

The spin-orbit geometry is conveniently considered in a
right-handed co-ordinate system defined by the angular-momentum
vectors, 
as illustrated in Fig.~\ref{fig1}.    The convention adopted here is
that the planetary-orbit and stellar-rotation 
angular-momentum vectors \bigo, \bigs\ lie in the $xy$ plane, with their vector sum
\bigL\
defining the $y$ axis.
Stellar rotation is assumed to be prograde (a choice that is
necessarily arbitrary, leading to ambiguities in several model
parameters; cf.\ the caption to Fig.~\ref{fig3}), and the inclination of the rotation axis to the line of
sight is required to be in the range $0 \le i_{\rm s} \le \pi/2$.
Orbital motion is retrograde with respect to the stellar rotation for
$\varphi_{\rm s} + 
\varphi_{\rm o} 
> \pi/2$, prograde otherwise
(where the $\varphi$ angles are defined in Fig.~\ref{fig1}).

Simple precession amounts to a rotation of \mbox{\bigo\ and \bigs} about
\bigL; observationally, this is equivalent to counter-rotation of the
line of sight.  Rather than impose a physical model of precession
(which requires assumptions about quantities such as the stellar
moment of inertia), in the present work the precessional angle $\psi$
is left as a free parameter for each epoch of observation.

\subsubsection{Stellar properties}

\citet{Barnes13} approximated the geometry of the rotationally distorted star
by an oblate spheroid; here,
the stellar surface is computed as a time-independent Roche
equipotential 
(cf., e.g., \citealt{Howarth01}), neglecting any gravitational
effects of the companion.  The global effective temperature is
defined 
by
\begin{align*}
\teff^4 = 
\int{\teffl^4 \dd A}
\left/ \int{\dc A} \right.
\end{align*}
where the integrations are over the distorted surface area, taking
into account gravity darkening.  For very rapid rotators this may not
correspond to any particular `observed' temperature (since the integrated line and
continuum spectra will not precisely match any single-star
standards),  but it is at least a well-defined quantity.

Values of $\teff = 3470$~K and polar gravity $\logg = 4.0$ (cgs) are
adopted here, following \citet{Briceno05} and \citet{Barnes13}.  These
values enter the analysis \textit{only} through the calculation of the
surface intensities, discussed below; otherwise, no assumptions are made, or 
are required, in respect of specific values of the mass or polar radius,
and other reasonable choices for $\logg$ would have no important
effect on the results.

\begin{figure}
\begin{center}
\includegraphics[angle=270, scale=0.35]{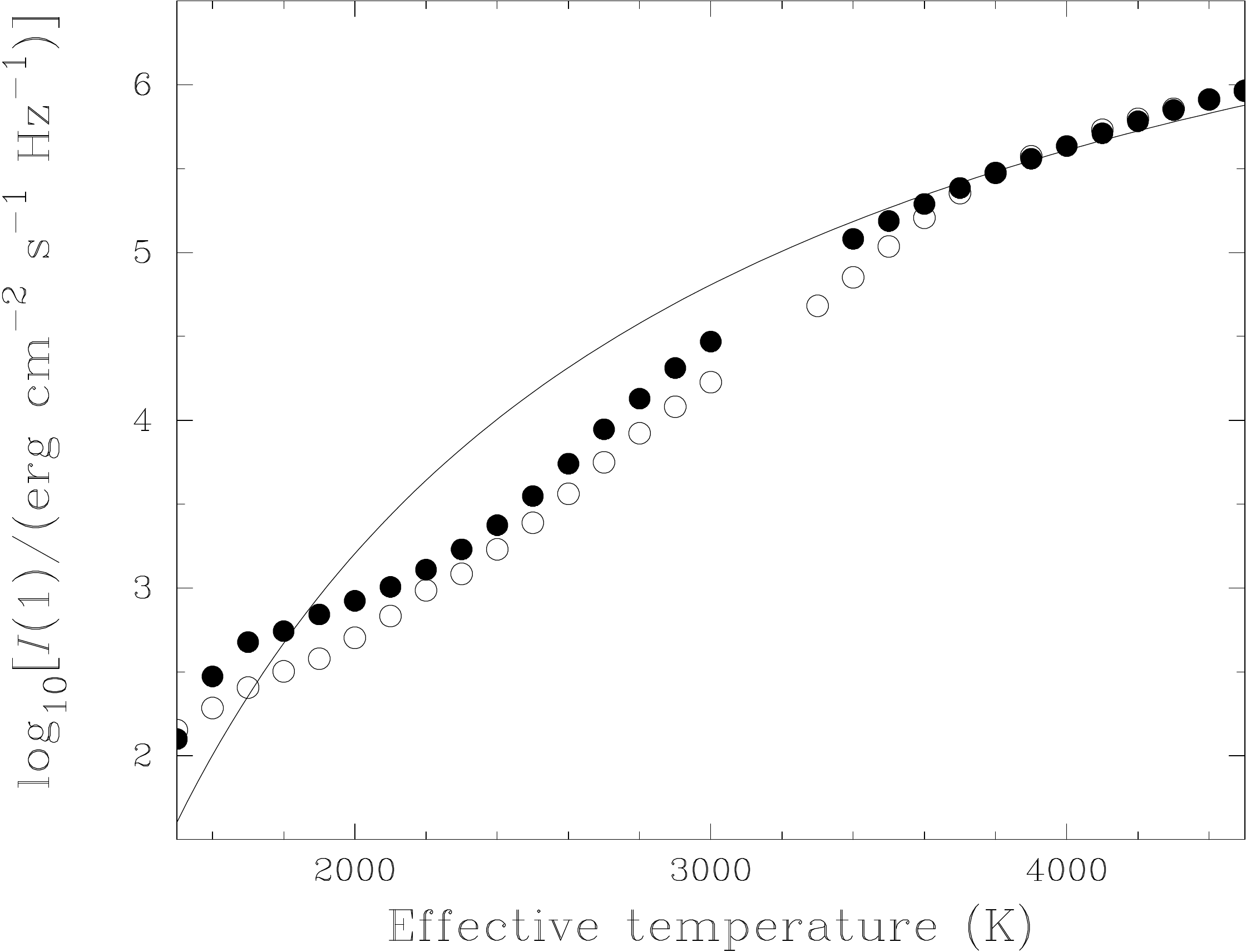} 
\caption{The $R$-band surface-normal intensity as a function of effective
  temperature.  Filled and open circles are 
solar-abundance model-atmosphere
results for $\logg = 4.5$ and 2.5~\mbox{dex cm s$^{-2}$},
respectively; the continuous curve is the (monochromatic) black-body result.
Note that the $y$ scale is logarithmic;  the model-atmosphere intensities can show large departures from black-body behaviour.}
\label{fig2}
\end{center}
\end{figure}

\subsubsection{Intensities}

Since the temperature and surface gravity must vary significantly over
the stellar surface (in order to generate the observed transit-curve
asymmetries), the dependence of the emergent intensity on these
quantities is of interest.  In this work, $R$-band surface intensities were
evaluated as $I(\mu,\teffl,g)$ by interpolation in the
`quasi-spherical' limb-darkening coefficients (LDCs) 
published
by \citet*{Claret12}, 
supplemented by surface-normal intensities
kindly provided by Antonio Claret.  
His 4-coefficient LDC
parametrization \citep{Claret00}, which reproduces the actual $I(\mu)$
distributions extremely well, was used.

The intensities derive from solar-abundance, line-blanketed,
non-LTE \textsc{Phoenix} model atmospheres (cf.\ \citealt{Claret12}).
Fig.~\ref{fig2} shows that the model-atmosphere emissivities
can depart substantially from black-body results, by up to a factor
$\sim$10 at 2.5~kK.  Thus although the principal intention of the
present analysis is to consider the consequences of an appropriate
treatment of gravity darkening, the use of model-atmosphere
intensities also represents a noteworthy if minor technical improvement
over previous work, which adopted black-body fluxes coupled to a
single, global, two-parameter limb-darkening law.

For the \citet{Barnes13} best-fit models, the implied equator--pole
temperature range is $\sim$3650--3350~K (for $\teff = 3470$~K); over
this temperature range, the model-atmosphere $R$-band surface-normal
intensity ratio is $\sim$0.47, while the black-body ratio is
$\sim$0.57.  Relaxing the assumption of black-body emission is
therefore liable to counteract the drive to larger values of \omomc\
required by adopting a smaller value for the $\beta$ exponent.

\begin{table}
\caption{Summary of Markov-chain Monte-Carlo results, 
for gravity darkening fixed at $\beta=0.08$.}
\begin{center}
\begin{tabular}{lcrcr}
\hline\hline
\multicolumn{1}{l}{Parameter} & &
\multicolumn{2}{c}{Distribution} &  \multicolumn{1}{c}{[Best]}           \\
\hline
%
%
$R_{\rm s}/a$                  && 0.518\pz  &  $^{+0.023 } / _{-0.031 }$ & 0.515\pz  \\
$R_{\rm p}/a$                  && 0.1050 &  $^{+0.0087} / _{-0.0094}$ & 0.1045 \\
\omomc\                          && 0.954\pz  &  $^{+0.012 } / _{-0.015 }$ & 0.953\pz  \\
$i_L (^\circ)$                 && 111.9  &  $^{+3.6   } / _{-4.8   }$ & 111.6  \\
$\varphi_{\rm o} (^\circ)$   && 0.8   &  $^{+2.0  } / _{-0.7  }$ & 0.2 \\
$\varphi_{\rm s} (^\circ)$   && 108.2  &  $^{+7.7   } / _{-4.9   }$ & 110.2  \\
$\psi (^\circ, 2009)$          && 272.1  &  $^{+2.7   } / _{-2.4   }$ & 272.1  \\
$\psi (^\circ, 2010)$          && 298.1  &  $^{+5.1   } / _{-4.3   }$ & 298.8  \\
\rule{0pt}{10pt}$\varphi_{\rm o} + \varphi_{\rm s} 
(^\circ)$   && 109.1  &  $^{+7.6   } / _{-4.5   }$ & 110.4  \\
$i_{\rm s} (^\circ, 2009)$   && 81.4   &  $^{+2.8   } / _{-3.2   }$ & 80.8   \\
$i_{\rm s} (^\circ, 2010)$   && 57.9   &  $^{+4.4   } / _{-4.9   }$ & 56.8   \\
$i_{\rm o} (^\circ, 2009)$   && 111.9  &  $^{+3.6   } / _{-4.8   }$ & 111.6  \\
$i_{\rm o} (^\circ, 2010)$   && 112.3  &  $^{+3.8   } / _{-4.8   }$ & 111.7  \\
\hline
\end{tabular}
\end{center}
  `Distribution' results are the median and 95\%\ confidence intervals
  (from $10^6$ MCMC replications), while the final column lists values
  for the individual trial model yielding the smallest rms
  residuals. $R_{\rm s}/a$ and $R_{\rm p}/a$ are the stellar polar
  radius and the planetary radius in units of the orbital semi-major
  axis; \omomc\ is the ratio of stellar angular rotation to the
  critical value.  Angles are defined in Fig.~\ref{fig1}; the sum
$\varphi_{\rm o} + \varphi_{\rm s}$, and the  
  stellar-rotation \& orbital inclinations, $i_{\rm s}$ \& $i_{\rm o}$,
are derived quantities, not free parameters in the model.
\label{t_results}
\end{table}

\begin{figure}
\begin{center}
\includegraphics[angle=0, scale=0.47]{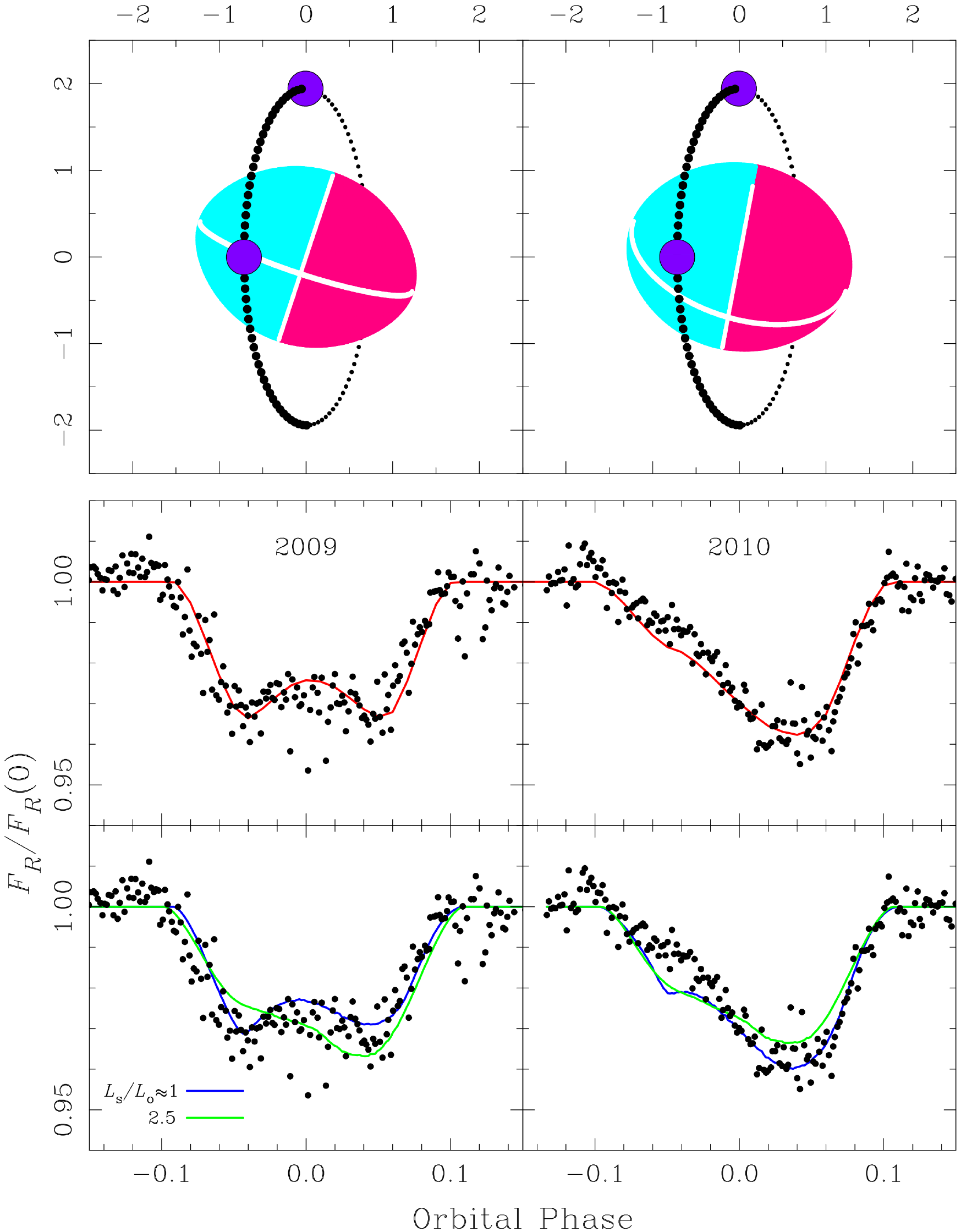} 
\caption{\textit{Upper panels:} system geometry at the two epochs, for
  the `best' solution summarized Table~\ref{t_results};  
the unit of length is the polar radius.  The colour coding of blue-
and red-shifted stellar hemispheres corresponds to prograde rotation;
retrograde rotation would give rise to identical light-curves, as
would mirror images of these panels.  The locations of the transiting
body at orbital phases 0.0 and 0.25 are shown, to indicate the
direction of orbital motion.  The projection of the total
angular-momentum vector onto the plane of the sky is aligned with the
$-x$ axis in each panel (and almost coincides with the orbital
angular-momentum vector).
\newline
\textit{Centre panels:} corresponding normalized 
model $R$-band light-curves and observations.
\newline
\textit{Bottom panels:}  the poorer `best-fit' models obtained with
angular-momentum ratios
$L_{\rm s}/L_{\rm o}$ constrained to values of 1 and 2.5 (q.v.\ $\S$\ref{sec:disco2}).  }
\label{fig3}
\end{center}
\end{figure}

\begin{figure}
\begin{center}
\includegraphics[angle=0, scale=0.7]{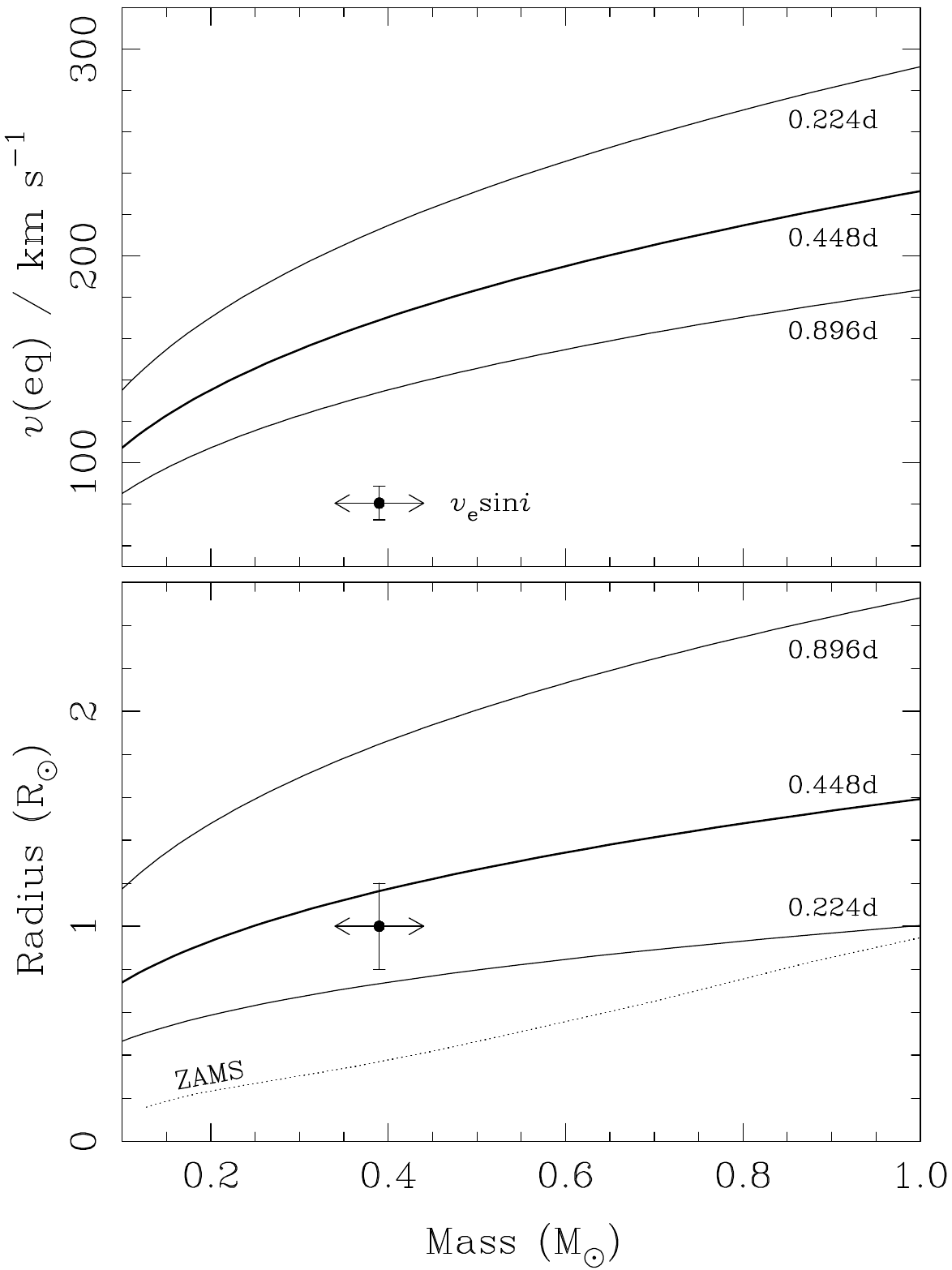} 
\caption{\textit{Upper panel:} 
Equatorial rotation velocity as a function of mass, for $\omomc =
0.954$ and three possible rotation periods.  The \vesini\ measurement
reported by \citet{vanEyken12} is shown (with its 1-$\sigma$ error)
at the mass range adopted by \citet{Barnes13}.\newline
\textit{Lower panel:} corresponding polar radii, together with the ZAMS
mass--radius
relationship (following \citealt{Eker15, Bertelli08}). The photometric
polar radius of $\sim 1.0 \pm 0.2$~\rsun\ is indicated ($\S$\ref{sec:disco}).}
\label{fig4}
\end{center}
\end{figure}

\section{Analysis}

\subsection{Observations}

\citet{vanEyken12} reported
$R$-band observations of 11 separate
transits in the 2009/10 observing season, and a further six in December
2010.  \citet{Barnes13} detrended and averaged these results to produce
mean `2009' and `2010' light-curves.  In order to approach as close
as reasonably possible a like-for-like comparison with their
results,
the \citeauthor{Barnes13} composite light-curves were digitized
and form the \mbox{basis} of the present analysis; the \citeauthor{Barnes13} ephemeris is
consequently also adopted, as a fixed quantity.   Because the dispersion
in the data appears not to be purely stochastic, all points were
equally weighted.

\subsection{Methodology}

In the model, the stellar geometry is determined by \omomc\ and by $R_{\rm s}/a$,
the polar radius expressed in units of the orbital
semi-major axis;
the exoplanet is characterized by its normalized
radius, $R_{\rm p}/a$.  Orbital/viewing geometry is defined by the angles
$\varphi_{\rm o}$, $\varphi_{\rm s}$, $i_L$, and $\psi$
(Fig.~\ref{fig1}).  The analysis requires all free parameters to be
the same at both epochs of observation, excepting the viewing angles
$\psi$.

Preliminary comparisons between the model and observations were
carried out by using a simple grid search, guided by the \citet{Barnes13}
results.  This pilot survey of parameter space was followed by a
Markov-chain Monte-Carlo (MCMC) analysis using a standard
Metropolis-Hastings algorithm (defaulting to $10^6$ replications and uniform
priors).

\subsection{Results}

A test run with $\beta = 0.25$ initially recovered
essentially the same geometry as found by \citet{Barnes13}, although
after $\sim 3 \times 10^5$ MCMC replications the fit 
migrated to an unphysical (though statistically marginally better)
solution, involving a grazing transit of a planet $\sim 5 \times$
larger than its parent star.

The $\beta = 0.08$ run did not suffer this problem, returning the
parameter set summarized in Table~\ref{t_results}.   Fig.~\ref{fig3}
\mbox{illustrates} the implied geometry, and confronts the predicted
light-curve with the data.  Disappointingly, the technical
improvements to the basic \citet{Barnes13} model implemented here lead
to a somewhat poorer overall match than they achieved.  In part, this
is a consequence of requiring a consistent parameter set at both
epochs (the two datasets can be matched extremely well if modelled
separately, as one might expect, given the number of free
parameters), but it is also suggestive of possible limitations  of the model.
%

\section{Discussion}
\label{sec:disco}

Phenomenologically, the solution obtained here provides a reasonably
satisfactory match between observed and predicted normalized
light-curves; however, it has \textit{physical} implications which
cast doubt on the completeness, or correctness, of the under\-pinning
model.

\subsection{Angular-momentum expectations}
\label{sec:disco1}

The magnitude of the stellar-rotation angular momentum 
for a star of mass $M_{\rm s}$ and polar radius $R_{\rm s}$ is
\begin{align*}
L_{\rm s} = I \omega 
\simeq \beta_{\rm g}^2 M_{\rm s} R_{\rm s}^2 \omega
\end{align*}
where $I$ is the moment of inertia, $\omega$ is the rotational
frequency, and $\beta_{\rm g}$ is the fractional radius of gyration.
A 
non-rotating 0.4\msun\ 
star approaching the zero-age main sequence has $\beta_{\rm g}^2 \simeq {0.19}$
\citep{Claret12a}, giving
\begin{align*}
\frac{L_{\rm s}}{\text{ kg m}^2\text{ s}^{-1}} = 
1.58 \times 10^{43}
\left[{ \frac{M_{\rm s}}{0.4\msun} }\right]^{3/2}
\left[{ \frac{R_{\rm s}}{\rsun} }\right]^{1/2} 
\left[{ \frac{\omega}{\omega_{\rm c}} }\right]
\left[{ \frac{\beta_{\rm g}^2}{0.19} }\right]
\end{align*}
where each bracketed term is intended to be of order unity (using
values
for mass
and radius based on discussions in \citealt{vanEyken12} and \citealt{Barnes13}).

The magnitude of the planetary-orbit angular momentum for a planet of mass
$M_{\rm p}$ with 
semi-major
axis $a$ and orbital 
frequency $\omega_{\rm orb} (= 2\pi/P_{\rm orb})$ 
is 
\begin{align*}
L_{\rm o} &= M_{\rm p} a^2 \omega_{\rm orb},\\
&\simeq M_{\rm p} M_{\rm s}^{2/3} P_{\rm orb}^{1/3} 
\left[{  \frac{G}{\sqrt{2\pi}} }\right]^{2/3}\text{ (for }M_{\rm p} \ll
M_{\rm s});
\end{align*}
numerically, for $P_{\rm orb} = 0.4484$d \citep{vanEyken12},
\begin{align*}
\frac{L_{\rm p}}{\text{ kg m}^2\text{ s}^{-1}} = 
1.47 \times 10^{42}
\left[{ \frac{M_{\rm p}}{3\mjup} }\right]
\left[{ \frac{M_{\rm s}}{0.4\msun} }\right]^{2/3}
\end{align*}
whence the rotational:orbital angular-momentum ratio is
\begin{align*}
\frac{L_{\rm s}}{L_{\rm o}} \simeq 10.7
\left[{ \frac{\beta_{\rm g}^2}{0.19} }\right]
\left[{ \frac{\omega}{\omega_{\rm c}} }\right]
\left[{ \frac{M_{\rm s}}{0.4\msun} }\right]^{5/6}
\left[{ \frac{R_{\rm s}}{\rsun} }\right]^{1/2} 
\left[{ \frac{3\mjup}{M_{\rm p}} }\right].
\end{align*}
The major source of uncertainty in this ratio is the planetary
mass, which is constrained only by the \citeauthor{vanEyken12} upper
limit
($M_{\rm p}\le[{5.5}\pm{1.4}]\mjup$), but
the bracketed terms are, cumulatively, unlikely to differ from unity
by more than perhaps a factor $\sim$3 or so.\footnote{The \citet{Barnes13}
`joint solution' (their Table~3) has 
${L_{\rm s}}/{L_{\rm o}} \simeq 2.5$, consistent with
their adoption of the Solar value for squared normalised radius of gyration (the
`moment of inertial coefficient' in their terminology), 
$\beta_{\rm g}^2 = 0.059$.}

\subsection{Angular-momentum results}
\label{sec:disco2}

The empirical results
summarized in Table~\ref{t_results},
obtained in the absence of any constraint on the
angular-momentum ratio, yield
\begin{align*}
\frac{L_{\rm s}}{L_{\rm o}}  
\left[{
\equiv {\frac{\sin(\varphi_{\rm p})}{\sin(\varphi_{\rm s})}}
}\right]
=
0.014^{ +0.036}_{-0.013}
\end{align*}
(median, 95\%\ confidence intervals).  This is discrepant, by almost
three orders of magnitude, with the prediction of
$\S$\ref{sec:disco1}; furthermore, the negligible
orbital precession implied by the small value 
of $\varphi_{\rm o}$ is inconsistent with the
absence of transits at some epochs (e.g., \citealt{Kamiaka15}).

Reasonably extensive sampling of parameter space, including several
tens of millions of MCMC replications starting from multiple initial
parameter sets, encourages the view that the solution summarized in
Table~\ref{t_results} locates the global minimum in $\chi^2$
hyperspace.  However (and particularly given that the model is constrained by
observations only two epochs), the question arises as to whether a
physically better model may exist with lower, but still acceptable,
statistical probability -- that is, does a preferable
solution occur at a local $\chi^2$ minimum?

To investigate this issue, further solutions were sought, again
through the MCMC process but imposing a variety of constraints on
the ${L_{\rm s}}/{L_{\rm o}}$ ratio.  In all these experiments, the angular-momentum
ratio was found always to drive towards the smallest allowed
values.   Figure~\ref{fig3} illustrates the outcomes of two such
experiments,  one in which ${L_{\rm s}}/{L_{\rm o}}$ was fixed at the 
\citeauthor{Barnes13} value of 2.5, and one in which it was required
to be $\ge$1 (with the outcome that the chain settled on a value very close to
1).  Neither of these models, nor any others examined, can be
considered as giving satisfactory fits.

\subsection{Consequences of stellar precession}

In the basic
\citet{Barnes13} model explored here,
a large
part of the light-curve variability between epochs arises through
precessional `nodding' of the star
(almost independently of the orbital angular-momentum issue discussed above).
This nodding gives rise to two
potentially observable diagnostics.  First, because of changes in
$\sin{i_{\rm s}}$, variability is expected in the projected equatorial
rotation velocity, \vesini\ (by a factor $\sim 1.2$ between the 2009
and 2010 epochs).  This may be easier to study spectroscopically than
the Rossiter--McLaughlin effect, because the variability timescale is
very much longer (allowing acquisition of better data).

Secondly, because the hotter polar regions of the star
are presented towards the observer in 2010, the system is predicted to
be brighter, by as much as $\Delta{R} = 0{\fm}30$ for the `best'
solution of Table~\ref{t_results}.
The PTFO photometry is in clear contradiction with this prediction.
Although there is significant non-orbital variability, the
observations shown by
\citet{vanEyken12} fall in the 
range $R \simeq 15.20 \pm 0.05$ at both 2009 and 2010 epochs (their
Figs.~2 and 3; the \textit{extreme} peak-to-peak range is only 0{\fm}17).  This is a
strong argument against the basic foundation of
the \citeauthor{Barnes13} model: any significant changes in the transit
morphology resulting from precession of a gravity-darkened
star are necessarily
accompanied by changes in the overall
brightness\footnote{The \citeauthor{Barnes13} $\beta=0.25$ solution implies
$\Delta{R} \simeq 0{\fm}2$.}
-- which is not
observed.

\subsection{Stellar rotation}

As anticipated, the solution with $\beta = 0.08$ requires a large (and
reasonably well-defined) value for \omomc.  The associated values of stellar
mass, radius, and equatorial rotation are not independent, but it is
straightforward to compute consistent sets of values for given \omomc\
and rotation period $P_{\rm rot}$. \citet{vanEyken12} found a signal
with $P = 0.448$~d in out-of-transit photometry, suggesting the
possibility of approximate rotational/orbital synchronization;
Fig.~\ref{fig4} illustrates the stellar equatorial rotation velocity
and polar radius as functions of mass for this $P_{\rm rot}$, and for 
values that are a factor two different in each direction.

For $V_0 \simeq 16.1$, $\teff \simeq 3.5$~kK, and $d \simeq
330$~pc \citep{Briceno05}, the effective stellar radius must be $\sim
1.1 \pm 0.2$~\rsun\ (polar radius $\sim 1.0 \pm 0.2$~\rsun),
as judged from \textsc{marcs} and
\textsc{atlas} model-atmosphere fluxes (\citealt{Gustafsson08,
Howarth11a};  see also \citealt{Barnes13}).
Supposing the stellar mass to be $\sim0.4 \pm 0.05$~\msun\
\citep{Briceno05, Barnes13}, rotation must indeed be close to, or
somewhat faster than,
synchronous to match this radius (Fig.~\ref{fig4}), which in turn
implies an equatorial rotation velocity $\veq \gtrsim 160$~\kms.

%

\citet{vanEyken12} report 
$\vesini = 80.6 \pm 8.1$~\kms\ from observations obtained in 2011 February.
If we suppose the inclination at that epoch to be close to the 2010
December value, then $\veq \simeq 95$~\kms.
Rapid rotation may lead to underestimation of \vesini\ (because a
consequence
of gravity
darkening is relatively low visibility of equatorial regions; cf.,
e.g., \citealt*{Townsend04}), but the discrepancy between observed and
expected equatorial velocities is too large to be explained by this effect.
Though less secure than the photometric constraint, this is therefore a further
source of
conflict between the model
and observations.

\section{Conclusion}

The ingenious `precession + gravity darkening' model proposed
by \citet{Barnes13} to interpret transit photometry of \ptfo\ has been
tested using a more appropriate characterization of gravity darkening,
along with with more sophisticated treatments of surface intensities and
stellar geometry.  

Although the normalized transit light-curves can still be adequately
reproduced by the model, the solution offered here has an
implausibly
small ratio of rotational to orbital angular momenta.
While other, physically more 
acceptable
solutions are not completely ruled
out, reasonably extensive exploration of parameter space has failed to
locate any such solution.

Independently of this issue, the adoption of a smaller
gravity-darkening exponent than previously assumed leads inexorably to
the requirement of near-critical stellar rotation.  Such rapid
rotation raises two further, and more general, difficulties for the
model.  First, given the `known' radius, the projected rotational
velocity is predicted to be approaching a factor two greater than
observed.  Secondly, a substantially gravity-darkened star must
exhibit significant photometric variability associated with precession
of the rotation axis; no such variability is observed.

Collectively, these results suggest that either the basic model omits
important physics, or that a conventional transiting exoplanet is not
the correct explanation for the fading events in the \ptfo\ system.

\section*{Acknowledgments}
I thank Antonio Claret for kindly providing absolute intensities to
supplement his published limb-darkening coefficients;  Ingo
Waldmann \& Michel Rieutord for enlightening discussions; and Jason Barnes \&
Julian van Eyken for helpful exchanges (the former both as
pre-submission correspondent and constructive referee).

\bibliographystyle{mnras}
\bibliography{IDH}

\label{lastpage}

\end{document}